\documentclass[12pt]{article}
\pdfoutput=1
\usepackage{amssymb,graphicx}
\usepackage{epstopdf}
\usepackage{amsmath,amsfonts}
\usepackage{epsfig}
\usepackage{textcomp}
\usepackage[space]{cite}
\usepackage[breaklinks]{hyperref}
\usepackage[english]{babel}

\def\e{{\rm e}}
\def\al{\alpha}
\def\tbeta{\tilde{\beta}}

\def\d{\partial}
\def\l{\left(}
\def\r{\right)}

\def\t0{\tilde{0}}
\def\ta{\tilde{a}}
\def\tb{\tilde{b}}
\def\tc{\tilde{c}}

\newcommand{\be}{\begin{equation}}
\newcommand{\ee}{\end{equation}}
\newcommand{\bea}{\begin{eqnarray}}
\newcommand{\eea}{\end{eqnarray}}
\newcommand{\bg}{\begin{gather}}
\newcommand{\eg}{\end{gather}}
\newcommand{\bseq}{\begin{subequations}}
\newcommand{\eseq}{\end{subequations}}

\newcommand{\talpha}{\tilde{\alpha}}

\begin{document}

\vspace{10pt}
\begin{center}
  {\LARGE \bf Stability of self-accelerating Universe in modified gravity with dynamical torsion: the case of small background torsion} \\
\vspace{20pt}
V.~Nikiforova$^a$\\
\vspace{15pt}
  $^a$\textit{Institute for Nuclear Research of
         the Russian Academy of Sciences,\\  60th October Anniversary
  Prospect 7a, 117312 Moscow, Russia
  }\\
    \end{center}
    \vspace{5pt}

\begin{abstract}
We consider the model of modified gravity with dynamical torsion. This model was found to have promising stability properties about various backgrounds. The model admits a self-accelerating solution. We have shown previously that if the parameters are adjusted in such a way that the torsion is much greater than the effective cosmological constant, the self-accelerating solution is unstable: there are exponentially growing modes. Here we study the scalar perturbations in the case when the torsion is of the order of the effective cosmological constant. We find that there are no exponential instabilities. \end{abstract}

\section{Introduction}
There have been numerous attempts to construct consistent IR-modified gravities (for reviews see Refs. \cite{m-1, m-2, m-3, m-4, m-5, m-6}), having in mind the explanation of the late-time acceleration of the Universe. Indeed, many theories of IR-modified gravity have self-accelerating solutions. The problems occur, however, when one considers the perturbations. The self-accelerating solutions are often unstable because of the ghost and/or gradient instabilities in the spectrum of the linearized perturbations. 

We focus here on gravities with dynamical torsion, which, apart from being promising candidates for consistent infrared modified gravity, represent non-trivial generalizations of General Relativity (GR) from the geometrical point of few. These gravities treat the connection and vierbein as independent dynamical variables. The connection is capable of propagating due to the terms in the Lagrangian which are quadratic in torsion and curvature.

We note in passing that these theories are often, but not exclusively, considered in the framework of Poincar\'e gauge gravities (PGT) \cite{book1, book2, book3}, where the connection field directly interacts with spin of matter. This interaction leads to strong constraints on the background connection in the Universe (see Ref. \cite{Ni} and references therein). Nevertheless, in this and previous papers we consider arbitrary values of the background connection, since the matter interaction with gravity may be the same as in GR.

We discuss here a particular model of modified gravity with dynamical torsion. This model was previously studied in Refs. \cite{33, 34, 45, NRR-2, N} and found to have nice stability properties. Namely, it was shown that ghosts, gradient instabilities and tachyons are absent in the Minkowski background (see also Refs. \cite{HS-3, Sezgin} where the most general parity-preserving quadratic 9-parameter Lagrangian was considered), de Sitter and anti-de Sitter spaces and arbitrary torsionless Einstein backgrounds of sufficiently small curvature \cite{33, 34}. 

The model also admits a self-accelerating solution, without an explicit cosmological constant term in the action \cite{NRR-2}. This solution has the FRW metric and the effective cosmological constant $\lambda$, 
which is due to the non-trivial connection. 

An important issue is the stability of small perturbations about this self-accelerating solution. In Ref. \cite{N}, we made use of the (3+1)-decomposition and started our analysis with the scalar sector of perturbations, having in mind that it is this sector that is usually the most dangerous from the viewpoint of instabilities. We found that there are two propagating degrees of freedom in the scalar sector, the same number as in Minkowski background, i.e., there are no Boulware--Deser \cite{B-D} modes. We derived the dispersion relations for these degrees of freedom in the limit of large background torsion, 
$$ f \gg \lambda \log^2A \;,$$
where $A$ is the initial amplitude of the linear perturbations and $f$ is the torsion strength. We showed that the self-accelerating solution is unstable, since there is at least one exponentially growing mode in the scalar sector \cite{N}.

In this paper we consider another limiting case, \be  f \sim \lambda \label{472} \;.\ee Note that the connection of the order of $\lambda$ directly interacting with spin is not in conflict with any experimental constraints (see Refs. \cite{Ni} and \cite{clear constraint}). Thus, we can treat the model under study as a Poincar\'e gauge gravity.

We derive the dispersion relations for the two propagating degrees of freedom in the scalar sector. We show that there are no exponentially growing modes in this case. 

This paper is a follow-up of Ref. \cite{N}, so we often refer to it. We present the model in Section 2. In Section 3 we recap the general treatment of Ref. \cite{N}: we write down the field equations, make (3+1)-decomposition of the linear perturbations and count propagating degrees of freedom in the scalar sector. Section 4 contains our main results. We derive the dispersion relations for scalar perturbations in the case \eqref{472} and show that there are no exponential instabilities. We discuss our results in Section 5.

\section{The Model} \label{sec:1}
We make use of the tetrad formalism and consider vierbein and connection as independent fields. We follow the notations of Refs. \cite{45, 33, 34, N, NRR-2} and denote the vierbein by $\varepsilon^i_\mu$ and connection by $A_{ij\mu} = -A_{ji\mu}$, where $\mu = (0,1,2,3)$ is the space-time index, and $i,j = (0,1,2,3)$ are the tangent space indices. The Lorentz indices are raised and lowered using the Minkowski metric $\eta_{ij}$, so we do not distinguish upper and lower tangent space indices in what follows, if this does not lead to an ambiguity. The signature of metric is $(-, +, +, +)$.

We consider the model with action
\be  S= \int~d^4x~\varepsilon L  \; , \;\;\; L= \frac{3}{2} ( \tilde{\alpha} F -
\alpha R) +
 c_3 F^{ij}F_{ij} + c_4 F^{ij}F_{ji} + c_5 F^2 + c_6 (\epsilon \cdot F)^2 \; , \label{all_2} \ee
where $ \alpha, \tilde{\alpha}, c_3, c_4, c_5, c_6$ are coupling constants, 
 $ \varepsilon \equiv det(\varepsilon^i_\mu) ; $
$F_{ijkl}$ is the curvature tensor constructed with the connection $A_{ij\mu}$,
\be
F_{ijkl} = \varepsilon^\mu_k \varepsilon^\nu_l ( \partial_\mu A_{ij\nu} - \partial_\nu A_{ij\mu} + A_{im\mu}A_{m j\nu} - A_{jm\mu}A_{m i\nu} ) \; ; \label{503}
\ee
\be F_{ij}=\eta^{kl}F_{ikjl}\;,\;\; F=\eta^{ij}F_{ij} \; , \;\; \epsilon \cdot F \equiv \epsilon^{ijkl}F_{ijkl} \nonumber \; ;\ee
$\epsilon_{ijkl}$ is the Levi-Civita symbol defined in such a way that $\epsilon^{0123} = - \epsilon_{0123} = 1$;
$R_{ijkl}$ is the Riemannian curvature tensor,
\be R_{ijkl}= \varepsilon^\mu_k \varepsilon^\nu_l ( \partial_\mu \omega_{ij\nu} - \partial_\nu
\omega_{ij\mu} + \omega_{im\mu}\omega_{mj\nu} - \omega_{jm\mu}\omega_{mi\nu} )\; ; \label{504}
\ee \be R_{ij}=\eta^{kl}R_{ikjl}\;,\;\; R=\eta^{ij}R_{ij}\; , \nonumber \ee
where $\omega_{ij\mu}$ is the Riemannian spin-connection:
\be  \omega_{ij\mu} = \frac{1}{2}( C_{ijk}-C_{jik} - C_{kij} )\varepsilon^k_\mu   \nonumber
\ee
with
\be
C_ {ijk} = \varepsilon_j^\mu \varepsilon_k^\nu (\d_\mu \varepsilon_{i\nu} - \d_\nu \varepsilon_{i\mu}) \; . \nonumber
\ee
The connection $A_{ij\mu}$ can be represented as a sum
\be
 A_{ij\mu}=\omega_{ij\mu}+K_{ij\mu}\; , \nonumber
\ee
where $K_{ij\mu}$ is the contorsion tensor.

In this paper the following conditions are imposed:
\begin{align}
& c_3 \neq c_4 \label{all_340} \;, \\
& |c_3| \sim |c_4| \sim |c_5| \label{501} \; \\
& c_3+c_4=-3c_5 \;, \label{all_19} \\
&\alpha < 0, \;\;\;\; \tilde{\alpha} > 0, \;\;\;\; c_5 < 0, \;\;\;\; c_6 > 0  \label{all_13} \; , \\
& c_5+16c_6 < 0 \; . \label{701}
\end{align}
The combination $(c_3 - c_4)$ appears repeatedly in our formulas, and our reasoning is valid, only if this combination is non-zero, see also \cite{N}. The conditions \eqref{all_19} and \eqref{all_13} ensure that there are no pathological degrees of freedom in the Minkowski background \cite{HS-3, Sezgin, 45, 33}. The condition \eqref{701} guarantees the existence of the self-accelerating solution that we consider in this paper (see \cite{NRR-2} for details). It was shown in Ref.~\cite{33} that the strength of the gravitational interaction between the energy-momentum tensors is governed by the parameters $\alpha$ and $\talpha$. We therefore assume that  \be  |\al| \sim \talpha \sim M_{Pl}^2 \;. \label{all_328}  \ee We also assume the validity of  \eqref{501} for convenience. 

It was found in Ref. \cite{33} that there are three propagating modes at the linearized level in the Minkowski background: the massless spin-2 mode, the massive spin-2 mode with mass
\be m^2=\frac{\talpha(\tilde{\alpha}-\alpha)}{2\alpha c_5} \label{all_80} \ee
 and the massive spin-0 mode with mass \be m^2_0=\frac{\tilde{\alpha}}{16c_6} \;. \label{all_90} \ee The perturbations about the Minkowski background are free of ghosts and tachyons. It was also shown that in the theory equipped with the cosmological constant, there are only healthy degrees of freedom in maximally symmetric backgrounds, as well as in torsionless Einstein backgrounds of sufficiently small curvature.

We found in Ref.~\cite{NRR-2} that the model \eqref{all_2} admits a self-accelerating cosmological solution with spatially flat metric,
\begin{align}  \varepsilon^{\t0}_0 = 1 \;,\;\;\;\; \varepsilon^{\ta }_b = e^{\lambda t} \delta^{\ta}_b
 \;,\quad
A_{\t0 \ta \tb}&=f\delta_{\ta\tb}   \;, \qquad
A_{\ta\tb\tc}=g\epsilon_{\ta\tb\tc}   \;,  \label{all_81} \\
  A_{\t0 \ta b}&=e^{\lambda t}f\delta_{\ta\tb} \;, \quad  A_{\ta\tb c}=e^{\lambda t}g\epsilon_{\ta\tb\tc} \;, \nonumber
 \end{align}
with time-independent $\lambda$, $f$ and $g$, where $a, \tilde{a}=(1,2,3)$, and tilde denotes tangent space indices. The parameters $\lambda$, $f$, $g$ and $\alpha$, $\talpha$, $c_5$, $c_6$ are related as follows\footnote{We correct a misprint in Ref. \cite{N} where the sign of $g^2$ is wrong in the third line in eq. \eqref{all_201}.},
\begin{align}
&c_6=\frac{\talpha\lambda (\talpha f+\al \lambda)}{16(\lambda^2-4f^2)(\talpha f^2-\talpha\lambda f - 2\al\lambda^2)} \;,  \nonumber
\\
& c_5=\frac{\talpha[2\talpha f^2+\lambda f\talpha+\lambda^2(\talpha-2\al)]}{4\lambda(\lambda+2f)(f^2\talpha-\lambda f\talpha-2\al\lambda^2)}  \;,   \label{all_201}   \\
& g^2=-\frac{2\alpha \lambda^2 - \talpha f^2 + \talpha \lambda f}{\talpha} \;. \nonumber
\end{align}

In Ref. \cite{N} we studied the small perturbations about the self-accelerating background \eqref{all_141} and considered the case of small $\lambda$ and large enough $f$. In that case eqs. \eqref{all_201} give
\be  c_6 \sim O(\lambda) \;, \quad  c_5 \sim O(\lambda^{-1}) \;, \quad g = \pm f+ O(\lambda) \;. \ee
We found that there are no Boulware--Deser modes, but some dynamical modes are exponentially growing, \be  \omega \sim -i\frac{f^2}{\lambda} \qquad \text{and} \qquad \omega \sim -i\sqrt{-\frac{\lambda^3f}{k^2}}  \;. \label{all_406} \ee The first instability can be pushed beyond the UV cutoff which is necessarily present in the theory and thus is not fatal, whereas the second one  makes the background \eqref{all_81} unstable.

In this paper we study small perturbations about the background \eqref{all_81} in another limiting case:
\be |f| = \delta\lambda \;, \qquad \text{where} \quad \delta \sim 1 \;. \label{600} \ee We recall that for positive $\lambda$ we have negative $f$ in accordance with \eqref{all_13} and the third formula in \eqref{all_201}, i.e. actually we have $$ f= -\delta \lambda \;.$$
In this case eqs. \eqref{all_201} give
\begin{align}
&c_6= \frac{\al\Xi(\Xi\delta+1)}{16\lambda^2(1-4\delta^2)(\Xi\delta^2 + \Xi\delta + 2)}\;,  \label{703}
\\
& c_5=\frac{\al\Xi(2\Xi\delta^2 - \Xi\delta + \Xi + 2)}{4\lambda^2(2\delta-1)(\Xi\delta^2+\Xi\delta + 2)}  \;,   \label{508}   \\
& g^2=\frac{\lambda^2(\Xi\delta^2+\Xi\delta+2)}{\Xi} \;, \label{706}
\end{align}
where $$ \Xi \equiv \frac{\talpha}{|\al|} \sim 1 $$
in accordance with \eqref{all_328}. 
In other words,
\be   |g| \sim \lambda \; , \quad \quad c_5 \sim c_6 \sim \frac{\talpha}{\lambda^2} \sim O(\lambda^{-2})\;.    \ee
We are going to see that the case \eqref{600} is stable.

Equations \eqref{703} - \eqref{706} show that the overall scale $|\al|$ enters the action \eqref{all_2} as a pre-factor. This implies that the equations for perturbations, written in terms of $f$, $g$ and $\lambda$, involve the ratio $\Xi =\talpha/|\al|$ and not $\talpha$ and $\al$ themselves. The same property holds for the dispersion relations.

Let us define the range of parameter $\delta$ such that the conditions \eqref{all_13} - \eqref{701} are satisfied. According to \eqref{all_13}, $c_6$ should be positive while $c_5$ should be negative. We see from eq. \eqref{703} that $c_6$ is positive when $1-4\delta^2<0$, i.e. \be \delta>\frac{1}{2} \;. \label{704} \ee
For $c_5$, the numerator in \eqref{508} is negative for any $\delta$,
$$  \al\Xi(2\Xi\delta^2 - \Xi\delta + \Xi + 2) = \al\Xi[\Xi(2\delta^2 - \delta + 1) + 2] < 0  \;,$$ because $ (2\delta^2 - \delta + 1)>0 $ and $\al <0$ by \eqref{all_13}. The denominator in \eqref{508} is positive if $\delta > 1/2$. Thus, $c_5$ is negative when \eqref{704} is satisfied. 
Finally, we substitute the eqs. \eqref{703} and \eqref{508} into the condition \eqref{701} and get the following inequality:
$$ c_5+16c_6 =  \frac{\al\Xi(2\Xi\delta^2 + \Xi\delta - \Xi + 2)}{4\lambda^2(2\delta+1)(\Xi\delta^2+\Xi\delta+2)} < 0\;,$$ which is sutisfied for $$ 2\Xi\delta^2 + \Xi\delta - \Xi + 2 = \Xi(2\delta^2 + \delta - 1)+2 >0 \;.  $$ The latter is automatically satisfied for $\delta$ restricted by \eqref{704}. Thus, the conditions \eqref{all_13}, \eqref{701} impose the only restriction \eqref{704} on $\delta$.

\section{Field equations}
In this section we review the analysis of Ref. \cite{N}, albeit briefly, for better understanding of Section 4.   
\subsection{Conformal transformation and field equations}
Starting from this moment, it is convenient to use the conformal time, \be
\eta = \int e^{-\lambda t}dt = -\frac{1}{\lambda}e^{-\lambda t} \;, \nonumber
\ee and change the variables:  
\be \varepsilon^i_\mu =
\e^{\phi}  e^i_\mu \;,\;\; \varepsilon_j^\nu =
\e^{-\phi}  e_j^\nu \;, \label{all_25} \ee
where
 \be \e^{\phi} = a(\eta) = -\frac{1}{\lambda \eta} \;.  \label{all_401} \ee
We do not make any scaling of connection, and keep the $A_{ij\mu}$ unchanged. Thus, the connection with all Lorentz indices now reads:
\be  A_{ijk} \equiv A_{ij\mu}e_k^{\mu} \;.\ee Upon the conformal transformation \eqref{all_25}, the background solution \eqref{all_81} can be written in terms of $e^i_{\mu}$ and $A_{ij\mu}$ as follows:
\be  e^i_\mu = \delta^i_\mu
 \;,\quad
   A_{\t0 \ta b}=e^{\phi}f\delta_{\ta b}   \;, \quad
  A_{\ta\tb c}=e^{\phi}g\varepsilon_{\ta\tb c}  \;.  \label{all_141}  \ee

Now we substitute \eqref{all_25} to eqs. \eqref{all_2} - \eqref{504} and derive the field equations. These are:

The gravitational equation:
\begin{align}
{\cal G}_{ij} \equiv \; & \frac{3}{2} \talpha \e^{-2\phi} \l F_{ij} - \frac{1}{2}
\eta_{ij}F  \r - \frac{3}{2} \alpha \e^{-2\phi} \l R_{ij} -
\frac{1}{2} \eta_{ij}R  \r
\nonumber\\
&+ \e^{-4\phi}c_3 \l F_{ki} F_{kj} + F_{kl} F_{kilj} \r + \e^{-4\phi}c_4 \l F_{ik}
F_{kj} + F_{lk} F_{kilj} \r + 2 \e^{-4\phi}c_5 F F_{ij}
\nonumber\\
&+ 2 \e^{-4\phi}c_6 \epsilon_{klmi} F_{klmj} (\epsilon\cdot F) -
\e^{-4\phi}\frac{1}{2}\eta_{ij} L^{(2)}  \nonumber \\
&  - \frac{3}{2} \alpha \e^{-2\phi} \left[ -2 e^\mu_i e^\nu_j
\nabla_\mu \nabla_\nu \phi - \eta_{ij} g^{\mu \nu} \nabla_\mu
\nabla_\nu \phi + 2  e^\mu_i e^\nu_j \d_\mu \phi \d_\nu \phi -2
\eta_{ij} g^{\mu \nu} \d_\mu \phi \d_\nu \phi \right. \nonumber
\\
& \left. - \eta_{ij} \l -3 g^{\mu \nu} \nabla_\mu \nabla_\nu \phi
-3 g^{\mu \nu} \d_\mu \phi \d_\nu \phi \r \right]  = 0  \;. \label{all_21}
\end{align}
Here 
\begin{align*}
L^{(2)} & =  c_3 F_{ij}F_{ij} + c_4 F_{ij}F_{ji} + c_5 F^2 + c_6 (\epsilon \cdot F)^2 
 \;;
\end{align*}
$F_{ijkl}$, $R_{ijkl}$ involve vierbein $e^{\mu}_i$ and connection $A_{ij\mu}$; $\nabla_{\mu}$ is the Riemannian covariant derivative constructed from $e_i^{\mu}$.

The torsion equation:
\begin{align}
{\cal T}_{ijk} \equiv \;& 
\e^{-3\phi}\left\{ \left[\eta_{ik} \l D_m P_{jm} - \frac{2}{3} D_j P\r -
D_i P_{jk}\right]  - \left[\eta_{jk} \l D_m P_{im} - \frac{2}{3}
D_i P\r - D_j P_{ik} \right]\right\}
\nonumber\\
& +4 \e^{-3\phi}c_6 \epsilon_{ijkm} D_m (\epsilon\cdot F) + \e^{-3\phi}S_{ijk} + \e^{-\phi}H_{ijk}   \nonumber \\
& + 3\talpha\e^{-\phi}\l    \eta_{ik} e^\mu_j \d_\mu \phi
- \eta_{jk} e^\mu_i \d_\mu \phi \r = 0 \;,
\label{all_22}
\end{align}
where $P_{ij}$ and $P$ are:
\begin{align*}   & P_{ij} = c_3F_{ij} + c_4 F_{ji}\;, \nonumber \\ & P=\eta^{ij}P_{ij} \;;   \end{align*}
the covariant derivative $D_i$ involves the vierbein $e^{\mu}_i$ and connection $A_{ij\mu}$,
\be  D_iB_j  \equiv  e^\mu_iD_\mu B_j = e^\mu_i( \partial_\mu B_{j} - A_{ lj\mu}B_{l} ) \;; \nonumber \ee
$S_{ijk}$ and $H_{ijk}$ are defined as follows: \begin{align*} S_{ijk} &= \frac{2}{3\talpha} H_{mnk} \l \eta_{im} P_{jn}
- \eta_{jm} P_{in} -\frac{2}{3} \eta_{im}\eta_{jn} P + 2c_6
\epsilon_{ijmn} (\epsilon \cdot F) \r \;,  \\ &\frac{2}{3\talpha} H_{ijk} = K_{ikj} -
K_{jki} - K_{ill} \eta_{jk} + K_{jll} \eta_{ik} \;. \label{all_41} \end{align*}
Note that because of invariance of the action under space-time gauge transformations and local Lorentz transformations, not all of the field equations are independent (see Sec.3 of \cite{N} for details).

\subsection{Perturbations about self-accelerating background}
Here we discuss the linear perturbations about the self-accelerating background \eqref{all_141}. Firstly, we make the 3-dimensional Fourier decomposition. We use the same notations for the Fourier-transformed variables and replace spatial derivatives $\d_{\ta} \equiv e_{\ta}^\mu \d_\mu$ with $ik_{\ta}$, where ${\bf k}$ is the 3-dimensional momentum. Secondly, we use $(3+1)$-decomposition of perturbations, since the background \eqref{all_141}  is invariant under spatial rotations. This means that we decompose any 3-dimensional tensor into its irreducible components with respect to the small group $ SO(2)$ of rotations around the spatial momentum, and obtain three independent sectors: scalar sector (helicity-0), vector sector (helicity-1) and tensor sector (helicity-2). These sectors can be considered separately, since the field equations are linear.

Known examples show that the most dangerous with regard to instabilities is the scalar sector of perturbations. Therefore, in this paper we are interested only in the scalar sector, leaving the vector and tensor sectors for future studies. 

The full contorsion tensor can be written as follows,
\be  K_{ij\mu}=A_{ij\mu \; (0)}+k_{ij\mu}  \;, \nonumber  \ee    where $A_{ij\mu \; (0)}$ is the background quantity \eqref{all_141} and $k_{ij\mu}$ is the first order perturbation. One decomposes the first order contorsion tensor $k_{ijk} = k_{ij\mu}\delta_k^{\mu}$ into its helicity components. 

Tensor $k_{ijk}=-k_{jik}$ contains helicity 0, 1, 2 components only. We concentrate here on the scalar sector (formed by the helicity-0 components):
\begin{align*}
k_{0a0} & = - k_{a00} = k_{a} \xi \;,
\\
k_{0ab} &= - k_{a0b} = k_{a} k_{b} \chi + \delta_{ab} \sigma +
\epsilon_{abc} k_{c} \rho \;,
\\
k_{ab0} &= \epsilon_{abc} k_{c} \theta \;,
\\
k_{abc} &= \epsilon_{abd} k_{c} k_{d} Q + (k_{a} \epsilon_{bcd} - k_{b}
\epsilon_{acd}) k_{d} u + (k_{a} \delta_{bc} - k_{b} \delta_{ac} ) M \;.
\end{align*}
There are 8 scalar components $\xi$, $\chi$, $\sigma$, $\rho$, $\theta$, $Q$, $u$ and $M$.

After conformal transformation for the vierbein \eqref{all_25} we have  \be e^i_\mu = \delta^i_\mu + \epsilon_{\mu}^i \nonumber  \;,  \ee
In what follows we use the standard gauge of the General Relativity - 
the conformal Newtonian gauge - which includes, inter alia, $$ e_{0 a}=0 \;.$$ We also impose the following gauge condition: $$ e_{\mu\nu}=e_{\nu\mu} \;.$$  After this gauge fixing we have 2 scalar components of tensor $\epsilon_{i\mu}$:
\begin{align*}
&\epsilon_{00}=-\Phi \;, \\
&\epsilon_{ab}=\Psi\delta_{ab} \;.
\end{align*}
Thus, in total we have 10 scalar variables. Using eqs.\eqref{all_21}-\eqref{all_22} and the above decomposition we can derive the field equations for these variables. These equations are written in Appendix A. We have designed a computer code to check the procedure of deriving equations and perform further manipulations with them. The code is available at \cite{code2}.

\subsection{Reducing the system of field equations}
In total, we have 14 field equations: 6 of them are second order in time derivatives and 8 are first order. The second order equations are eqs.~\eqref{c5}, \eqref{c9}, \eqref{c10}, \eqref{c11}, \eqref{c13} and \eqref{c14}, while the first order equations are \eqref{c1} - \eqref{c4}, \eqref{c6} - \eqref{c8} and \eqref{c12}. 
These 14 equations are not totally independent. There are 4 Bianchi identities relating field equations with each other (see Sec. 6.2 of \cite{N}). These 4 identities can be used to express eqs.~\eqref{c5}, \eqref{c9}, \eqref{c13} and \eqref{c14} (which are second order) in terms of other components. Therefore, these equations can be ignored, and from this moment we consider the system of 10 equations: 8 first order equations \eqref{c1}-\eqref{c4}, \eqref{c6}-\eqref{c8}, \eqref{c12} and 2 second order equations \eqref{c10}, \eqref{c11}. 

In what follows it is convenient to use the effective values of $f$, $g$, $\lambda$, $\al$ and $\talpha$:
\begin{align} f, \; g, \; \lambda &\equiv {\mathfrak f}e^{-\phi},\; {\mathfrak g}e^{-\phi},\; \Lambda e^{-\phi} \;, \nonumber \\
 \al, \; \talpha &\equiv \beta e^{-2\phi},\; \tbeta e^{-2\phi} \label{507} \;. \end{align}
The relations between ${\mathfrak g}$, $c_5$, $c_6$ and $\Lambda$, ${\mathfrak f}$, $\beta$, $\tbeta$ have the same form as in eqs.~\eqref{703} - \eqref{706}, with the substitution $$ f \longrightarrow {\mathfrak f}\;, \quad \alpha \longrightarrow \beta \;, \quad \text{etc.}$$ In the manipulations described in this section we use the relations \eqref{703} - \eqref{706} to simplify the equations.

The two second order equations \eqref{c10} and \eqref{c11} can be replaced by first order equations. To this end, we combine these second order equations with the remaining first order equations and their time derivatives. The resulting equations \eqref{d1}, \eqref{d2} are first order, their explicit forms are given in Appendix B.  Therefore, at this stage we have 10 first order equations, \eqref{c1} - \eqref{c4}, \eqref{c6} - \eqref{c8}, \eqref{c12}, \eqref{d1}, \eqref{d2}, for 10 variables describing scalar perturbations.

The variables in the resulting system of equations are of two types. The variables of the first group, \be \sigma\;, \; \Phi\;,\; \xi\;, \; \theta \;, \label{all_300} \ee enter these equations without time derivatives. The variables of the second group, \be \Psi\;, \;\chi\;, \;\rho\;, \;Q\;, \;u\;, \; M \;,  \label{all_86} \ee enter with first time derivatives. One can make use of any 4 of the remaining 10 equations to express the 4  variables \eqref{all_300} in terms of the variables \eqref{all_86} and their first time derivatives. There remain 6 first order equations for the 6 variables \eqref{all_86}.

Moreover, two linear combinations of the 10 equations \eqref{c1} - \eqref{c4}, \eqref{c6} - \eqref{c8}, \eqref{c12}, \eqref{d1}, \eqref{d2} are actually algebraic and involve only the variables \eqref{all_86} (for details, see Sec. 6.3 of \cite{N}). Therefore, one can make use of these two algebraic equations to express 2 variables from the set \eqref{all_86} ($M$ and $\rho$) in terms of four remaining variables from this set ($\Psi$, $\chi$, $Q$, $u$). After this is done, we have four first order equations for the 4 variables from \eqref{all_86}.

Thus, we have 2 degrees of freedom. All the procedure is done by making use of the code \cite{code2} and until now is totally equivalent to the one described in \cite{N}.

\section{Limit of small $\Lambda$}
In practice, it is convenient to express $\sigma$, $\xi$, $\Phi$, $\theta$ by making use of eqs.~\eqref{c3}, \eqref{c4}, \eqref{c7} and \eqref{c8}, respectively, and substitute these into eqs. \eqref{c2}, \eqref{c6}, \eqref{d1} and \eqref{d2}. After that we express $M$, $\rho$, $M^\prime$ and $\rho^\prime$ from the algebraic equations and their time derivatives, and substitute these in eqs. \eqref{c2}, \eqref{c6}, \eqref{d1}, \eqref{d2}. Thus, our linearized system reduces to four first order equations \eqref{c2}, \eqref{c6}, \eqref{d1}, \eqref{d2} written in terms of variables $\Psi$, $\chi$, $Q$, $u$.

Let us now consider the case of $\Lambda$ being the smallest parameter of the problem and  $$ {\mathfrak f}=-\delta\Lambda \;, \qquad \delta \sim 1 \;.$$ In what follows we make use of the relations \eqref{703} - \eqref{706} (written for the effective values ${\mathfrak f}$, ${\mathfrak g}$ etc.), and obtain equations in terms of the time-dependent parameter $\Lambda$ and time-independent parameters $\delta$, $\Xi$ and also $b \equiv \frac{c_3}{c_4}$ (the latter ratio, in fact, does not enter the final result).  Then, as long as we are working in the small-$\Lambda$ limit, the first order equations for $\Psi$, $\chi$, $Q$, $u$ become series in $\Lambda$.

For small $\Lambda$, parameters ${\mathfrak f}$, ${\mathfrak g}$, $\Lambda$, $\beta$, $\tbeta$ have slow dependence on time. We seek for solutions in the WKB form $F \sim e^{i\int\omega d\eta}$, where $F$ is any of the
variables $\Psi$, $\chi$, $Q$, $u$, and
\be
\omega \gg \Lambda, \quad k^2 \equiv k_ak_a \gg \Lambda^2 \;.  \label{all_103}
\ee
Then the time derivative is replaced by
\be
\d_0  \rightarrow i\omega \;.  \nonumber
\ee
In this way we obtain the system of 4 linear homogeneous algebraic equations \eqref{c2}, \eqref{c6}, \eqref{d1}, \eqref{d2} for $\Psi$, $\chi$, $Q$, $u$, which determine the dispersion relations $\omega_i=\omega_i(k), i=1,\dots,4$. Equating the determinant of the system to zero we get a fourth order equation for $\omega$. To the leading order in $\Lambda$ it reads:
\begin{align}  & [\omega^4 -(2\delta +1)k^2\omega^2 +4ik^2\delta(2\delta+1)\Lambda\omega- (2\delta + 1)(4\delta+1)k^2\Lambda^2]  \nonumber \\
& + K(\Xi, \delta, b)\Lambda\left( \frac{\Lambda}{k}\right)^{2n}\omega^3  =0 \;,  \label{6}  \end{align}
where $n \geq 0$, and $K(\Xi, \delta, b)$ is some coefficient. The particular structure of the term with $K(\Xi, \delta, b)$ is due to the cancellation that occurs in the term with $\omega^3$. Without the cancellation, this term would be proportional to $\omega^3k^2/\Lambda$. We have found by direct calculation that the cancellation occurs at least in two leading orders in $\Lambda$. We have not calculated the term with $K(\Xi, \delta, b)$, since it is subdominant in all regimes anyway, as we see below.  

Equation \eqref{6} is fourth order in $\omega$. Our approach is to find all 4 roots of eq.~\eqref{6} and then figure out the roots obeying $\omega \gg \Lambda$. In this way we gain confidence that no relevant solutions to eq.~\eqref{6} are lost. We are going to see that there are four roots of the dispersion equation \eqref{6}: two imaginary,
\be \omega \sim \pm i\Lambda \;, \label{702}\ee
and two real,
\be \omega \sim k \;.\label{744} \ee
We emphasize that the roots \eqref{702} do not obey $\omega \gg \Lambda$, which is our original assumption, see \eqref{all_103}. So, we cannot be confident of the instability with the time scale of order $\Lambda^{-1}$, but even if it is there, this time scale is large and hence the instability is not dangerous. To get to the result \eqref{702}, \eqref{744}, we consider limiting cases.
  
\begin{itemize}
\item $\omega \ll \Lambda$. In this case the term with $\omega=0$ is larger than others, and eq. \eqref{6} reduces to $$ (2\delta + 1)k^2\Lambda^2 =0 \;.$$ There are no solutions.

\item $\omega \sim \Lambda$. The second, third and fourth terms in square brackets survive and we have $$ -(2\delta +1)k^2\omega^2 +4ik^2\delta(2\delta+1)\Lambda\omega- (2\delta + 1)(4\delta+1)k^2\Lambda^2  \sim k^2\Lambda^{2}\;,$$ while the term with $K(\Xi, \delta, b)$ is smaller, $$ K(\Xi, \delta, b)\Lambda\left(\frac{\Lambda}{k}\right)^{2n}\omega^3 \sim  \left (\frac{\Lambda}{k}\right)^{2n}\Lambda^{4} \;. $$ Thus, we have in this case the second order equation \be  -(2\delta +1)k^2\omega^2 +4ik^2\delta(2\delta+1)\Lambda\omega- (2\delta + 1)(4\delta+1)k^2\Lambda^2 = 0 \;, \ee which has two roots (cf. eq. \eqref{702}): \begin{align} &\omega_{1} =  i\Lambda \;, \label{502} \\
&\omega_2=-i\Lambda(4\delta+1) \;. \label{721}
 \end{align} As we discussed above, these roots  do not obey $\omega \gg \Lambda$ and thus are irrelevant.
 
\item $\Lambda \ll \omega \ll k.$ The second term in the square brackets is the largest. The term with $K(\Xi, \delta, b)$ is again irrelevant: $$  \omega^3\Lambda\left( \frac{\Lambda}{k} \right)^{2n}  \ll  k^2\omega^2     \;. $$ Thus, eq. \eqref{6} reduces to $$ (2\delta + 1)k^2\omega^2 = 0  \;.$$ There are no roots.

\item $\omega \sim k.$ The terms with $\omega^4$ and $\omega^2$ survive, and we have the following equation: \be  \omega^4 -(2\delta +1)k^2\omega^2=0  \;. \ee  There are two roots, \be  \omega_{3,4} = \pm \sqrt{(2\delta+1)k^2}  \;. \label{748} \ee Recall that $\delta>0$, so what these frequencies are real. 

\item $\omega \gg k$. The term with $\omega^4$ is much larger than others, thus there are no solutions.

So, the four roots of eq. \eqref{6} are given by \eqref{502}, \eqref{721} and \eqref{748}, as promised. There is no instability with the time scale exceeding $\Lambda^{-1}$.

\end{itemize}

\section{Discussion}
In this paper we continued our analysis started in Ref. \cite{N}. We studied the stability of the self-accelerating solution \eqref{all_81} in the model \eqref{all_2} at the linearized level in the case of small background torsion \eqref{600}. We made use of the $(3+1)$-decomposition and considered the scalar sector of perturbations which has two degrees of freedom \cite{N}. In the case of small background torsion we have found two oscillating modes with frequencies $$ \omega= \pm \sqrt{(2\delta+1)k^2} \;.$$  Two other modes with the time scale of order $\lambda^{-1}$ may exist, but they are not covered by our analysis. These modes, even if naively unstable, correspond to the large time scale and hence are not dangerous. 

Our analysis reveals the following subtle feature of the model under study. In our case of small background torsion, 
$|f|, |g| \sim \lambda$ , 
the self-accelerating solution \eqref{all_81} tends to the Minkowski space as $\lambda \rightarrow 0$: 
$$  \varepsilon^{i }_{\mu} \rightarrow  \delta^{i}_{\mu}
 \;,\quad
A_{ij\mu} \rightarrow 0    \;.$$
However, in terms of the dynamics of perturbations in the gravitational sector, the Minkowski space and self-accelerating solution are different branches not connected to each other. Although the number of degrees of freedom in self-accelerating background \eqref{all_81} is the same as in Minkowski background, the dispersion relations are fundamentally different. 

We also note that the dispersion relations \eqref{502}, \eqref{721} and \eqref{748} are completely different from ones found in Ref. \cite{N} where the limit of large background torsion was considered (cf. \eqref{all_406}). Namely, the results \eqref{502}, \eqref{721}, \eqref{748} cannot be obtained by taking the limit $f \rightarrow \lambda$ in formulas of Ref. \cite{N}. This is not surprising. Indeed, in Ref. \cite{N} we examined the small perturbations in two regimes,
$$ k^2 \ll \lambda f \quad \text{and}   \quad  f^3\lambda \ll k^2 \ll \frac{f^5}{\lambda} \;,      $$
which are both inconsistent with our case $f \sim \lambda$ and $k \gg \lambda$.

Although it is likely that the modes \eqref{748} are healthy, we cannot rule out at the moment the possibility that these are ghosts. Also, to fully investigate the stability of the self-accelerating background in this model, the vector and tensor sectors need to be considered. We plan to study these issues in future. 

\section*{Acknowledgements}
The author is deeply indebted to Valery Rubakov for numerous discussions and suggestions. This work has been supported by Russian Science Foundation grant 14-22-00161.

\newpage

\section*{Appendix A}
Here we write the 14 scalar field equations. We use the notation \eqref{507} and recall the relation \eqref{all_19}.
Equations obtained from \eqref{all_21} are as follows.

$(00)$-component:
\begin{align}
{\cal G}^{(00)} \equiv & -96c_6k^2\Lambda {\mathfrak g}u^\prime - 48c_6k^2\Lambda {\mathfrak g} Q^\prime + (-9\tilde{\beta}{\mathfrak f} - 9\beta \Lambda - 576c_6{\mathfrak g}^2{\mathfrak f}) \Psi^\prime     \nonumber
\\
& + (576c_6{\mathfrak g}^2{\mathfrak f} + 9\tilde{\beta}{\mathfrak f})\sigma + (144c_6{\mathfrak g}^2\Lambda^2 + 9\beta\Lambda^2)\Phi \nonumber \\
& + [k^2(3\tilde{\beta} - 3\beta) - 9\tilde{\beta}({\mathfrak f}^2-{\mathfrak g}^2) + 144c_6{\mathfrak g}^2(\Lambda^2-8{\mathfrak f}^2)]\Psi     \nonumber
\\
& + (-6k^2\tilde{\beta}{\mathfrak g} + 384k^2c_6{\mathfrak f}^2{\mathfrak g})u + \frac{1}{2} (-6k^2\tilde{\beta}{\mathfrak g} + 384k^2c_6{\mathfrak f}^2{\mathfrak g})Q + 3ik^2\tilde{\beta}M    \nonumber
\\
& + (3k^2\tilde{\beta}{\mathfrak f} + 192k^2c_6{\mathfrak g}^2{\mathfrak f})\chi + 192ik^2c_6{\mathfrak f}{\mathfrak g}\rho + 48ik^2c_6\Lambda {\mathfrak g}\theta = 0
\tag{a1}  \label{c1}
\end{align}

$(a0)$-component:
\begin{align}
{\cal G}^{(a0)} \equiv & -i(3\tilde{\beta} + 4\Lambda {\mathfrak f} c_3)M^\prime -i [2\Lambda {\mathfrak g}(c_4-c_3) - 96c_6{\mathfrak g}(\Lambda-2{\mathfrak f})]\rho^\prime + (3\beta - 3\tilde{\beta})\Psi^\prime    \nonumber
\\
& + [-6{\mathfrak g}^2\Lambda c_5 + 4\Lambda c_3({\mathfrak f}^2-{\mathfrak g}^2)]\Psi     -i [6{\mathfrak g}^2\Lambda c_5 - 4\Lambda({\mathfrak f}^2-{\mathfrak g}^2)c_3]M \nonumber
\\
& + [-3\beta\Lambda - 3\tilde{\beta}{\mathfrak f} + 96c_6{\mathfrak g}^2(\Lambda - 2{\mathfrak f}) + 4\Lambda({\mathfrak g}^2-{\mathfrak f}^2)c_3 + 6\Lambda {\mathfrak g}^2c_5]\Phi \nonumber
\\
& - 4\Lambda {\mathfrak f}c_3\sigma -i (8\Lambda {\mathfrak f}{\mathfrak g}c_3 + 6\Lambda {\mathfrak f}{\mathfrak g}c_5)\rho + (6\Lambda {\mathfrak g}k^2c_5 + 4\Lambda {\mathfrak g}k^2c_3)u      \nonumber
 \\
& -i [4\Lambda({\mathfrak f}^2-{\mathfrak g}^2)c_3 - 6{\mathfrak g}^2\Lambda c_5 - 96c_6\Lambda(\Lambda - 2{\mathfrak f}) + 3\tilde{\beta}{\mathfrak f}]\xi    \nonumber
  \\
& -i [-8\Lambda {\mathfrak f}{\mathfrak g}c_3 - 6\Lambda {\mathfrak f}{\mathfrak g}c_5 - 96c_6{\mathfrak f}{\mathfrak g}(\Lambda-2{\mathfrak f}) - 3\tilde{\beta}{\mathfrak g}]\theta=0
\tag{a2}   \label{c2}
\end{align}

$(0a)$-component:
\begin{align}
{\cal G}^{(0a)} \equiv & (3\beta - 3\tilde{\beta})\Psi^\prime- 4ic_3({\mathfrak f}^2-{\mathfrak g}^2)M^\prime - 4i(c_4-c_3){\mathfrak f}{\mathfrak g}\rho^\prime    \nonumber
\\
& + i(12c_3{\mathfrak g}^2{\mathfrak f} + 12{\mathfrak g}^2{\mathfrak f}c_5 - 4c_3{\mathfrak f}^3)\xi -i (-12c_3{\mathfrak g}{\mathfrak f}^2 - 12c_5{\mathfrak g}{\mathfrak f}^2 + 4c_3{\mathfrak g}^3)\theta  \nonumber
\\
& -i [3\tilde{\beta}{\mathfrak f} - 96c_6{\mathfrak g}^2(\Lambda - 2{\mathfrak f}) + 12c_3{\mathfrak g}^2{\mathfrak f} + 12c_5{\mathfrak g}^2{\mathfrak f} - 4c_3{\mathfrak f}^3]M    \nonumber
\\
& -i [-3\tilde{\beta}{\mathfrak g} - 96c_6{\mathfrak g}{\mathfrak f}(\Lambda - 2{\mathfrak f}) + 12c_3{\mathfrak g}{\mathfrak f}^2 + 12c_5{\mathfrak g}{\mathfrak f}^2 - 4c_3{\mathfrak g}^3]\rho    \nonumber
\\
& + [-96c_6k^2{\mathfrak g}(\Lambda - 2{\mathfrak f}) - 4(c_4-c_3){\mathfrak g}{\mathfrak f}k^2]u + [3\tilde{\beta} - 4c_3({\mathfrak f}^2-{\mathfrak g}^2)]\sigma    \nonumber
\\
& + [-3\tilde{\beta}{\mathfrak f} + 96c_6{\mathfrak g}^2(\Lambda - 2{\mathfrak f}) - 12c_3{\mathfrak g}^2{\mathfrak f} - 12c_5{\mathfrak g}^2{\mathfrak f} + 4c_3{\mathfrak f}^3]\Psi   \nonumber
\\
& + [-3\beta\Lambda - 4c_3{\mathfrak f}({\mathfrak f}^2-{\mathfrak g}^2) - 4{\mathfrak g}^2{\mathfrak f}(c_4-c_3)]\Phi = 0
\tag{a3}    \label{c3}
\end{align}

$(ab)$-component, $k_{\ta}k_{b}$:
\begin{align}
{\cal G}^{(k\otimes k)} \equiv & -(48c_6\Lambda {\mathfrak g} - 96c_6{\mathfrak f}{\mathfrak g})Q^\prime - [-3c_5\Lambda {\mathfrak f} - \frac{3}{2}\tilde{\beta} - 3c_5({\mathfrak f}^2-{\mathfrak g}^2)]\chi^\prime - (96c_6{\mathfrak f}{\mathfrak g}-48c_6\Lambda {\mathfrak g})u^\prime   \nonumber
\\
& - [-\frac{3}{2}\beta + \frac{3}{2}\tilde{\beta} + 3c_5({\mathfrak g}^2-{\mathfrak f}^2-\Lambda {\mathfrak f})]\Psi - [3c_5\Lambda {\mathfrak f} + \frac{3}{2}\tilde{\beta} - \frac{3}{2}\beta + 3c_5({\mathfrak f}^2-{\mathfrak g}^2)]\Phi   \nonumber
\\
 & - (-48ic_6\Lambda {\mathfrak g} + 96ic_6{\mathfrak f}{\mathfrak g})\theta - (-48ic_6\Lambda {\mathfrak g} + 96ic_6{\mathfrak f}{\mathfrak g})\rho - [3ic_5({\mathfrak f}^2-{\mathfrak g}^2+\Lambda {\mathfrak f}) + \frac{3}{2}i\tilde{\beta}]\xi   \nonumber
 \\
& - (-3c_5{\mathfrak f}^3 - 3c_5\Lambda {\mathfrak f}^2 - 48c_6{\mathfrak g}^2\Lambda + 96c_6{\mathfrak g}^2{\mathfrak f} + 3c_5{\mathfrak g}^2{\mathfrak f} + \frac{3}{2}{\mathfrak f}\tilde{\beta})\chi    \nonumber
\\
& - (3c_5{\mathfrak f}^2{\mathfrak g} + 3c_5\Lambda {\mathfrak f}{\mathfrak g} - 3c_5{\mathfrak g}^3 - 48c_6\Lambda {\mathfrak f}{\mathfrak g} + 96c_6{\mathfrak f}^2{\mathfrak g} - \frac{3}{2}{\mathfrak g}\tilde{\beta})Q   \nonumber
\\
& - [-3ic_5({\mathfrak f}^2-{\mathfrak g}^2+\Lambda {\mathfrak f}) + \frac{3}{2}i\tilde{\beta}]M   \nonumber
\\
& - u(-96c_6{\mathfrak g}{\mathfrak f}^2 - 3c_5{\mathfrak f}^2{\mathfrak g} + 3c_5{\mathfrak g}^3 + 48c_6\Lambda {\mathfrak f}{\mathfrak g} - 3c_5\Lambda {\mathfrak f}{\mathfrak g} + \frac{3}{2}{\mathfrak g}\tilde{\beta})=0
\tag{a4}    \label{c4}
\end{align}

$(ab)$-component, $\delta_{ab}$:
\begin{align}
{\cal G}^{(\delta)} \equiv & (-3\tilde{\beta} +3\beta)\Psi^{\prime\prime} + (-16c_6\Lambda {\mathfrak g}k^2 - 32k^2c_6{\mathfrak f}{\mathfrak g})u^\prime + (-32c_6\Lambda {\mathfrak g}k^2 + 32k^2c_6{\mathfrak f}{\mathfrak g})Q^\prime    \nonumber
 \\
 & + (-192c_6{\mathfrak f}{\mathfrak g}^2 + 6\beta\Lambda + 3{\mathfrak f}\tilde{\beta})\Psi^\prime + (c_5k^2\Lambda {\mathfrak f} + \frac{3}{2}k^2\tilde{\beta} - c_5k^2{\mathfrak g}^2 + c_5k^2{\mathfrak f}^2)\chi^\prime + 3\tilde{\beta}\sigma^\prime    \nonumber
 \\
& + [ -ic_5k^2({\mathfrak f}^2-{\mathfrak g}^2+\Lambda {\mathfrak f}) -\frac{3}{2}ik^2\tilde{\beta}]\xi + (32ik^2c_6{\mathfrak f}{\mathfrak g} + 16ik^2c_6\Lambda {\mathfrak g})\rho + (32ik^2c_6\Lambda {\mathfrak g}     \nonumber
 \\
& + [-c_5k^2({\mathfrak f}^2-{\mathfrak g}^2+\Lambda {\mathfrak f}) + 48c_6\Lambda^2{\mathfrak g}^2 - 3\tilde{\beta}\Lambda {\mathfrak f} - 9\beta\Lambda^2 - \frac{3}{2}k^2\tilde{\beta} + \frac{3}{2}k^2\beta]\Phi    \nonumber
 \\
& + [-3\tilde{\beta}\Lambda {\mathfrak f} + 48c_6\Lambda^2{\mathfrak g}^2 + c_5k^2({\mathfrak f}^2-{\mathfrak g}^2+\Lambda {\mathfrak f}) - 384c_6{\mathfrak f}^2{\mathfrak g}^2 + \frac{3}{2}k^2\beta  \nonumber
\\
&+ 3\tilde{\beta}{\mathfrak f}^2 - \frac{3}{2}k^2\tilde{\beta} - 3\tilde{\beta}{\mathfrak g}^2]\Psi  - 32ik^2c_6 {\mathfrak f}{\mathfrak g})\theta  + (-3\tilde{\beta}{\mathfrak f} + 192c_6{\mathfrak g}^2{\mathfrak f})\sigma     \nonumber
\\
 & + (c_5k^2\Lambda {\mathfrak f}^2 - c_5k^2{\mathfrak g}^2{\mathfrak f} - \frac{3}{2}k^2{\mathfrak f}\tilde{\beta} + c_5k^2{\mathfrak f}^3 +32k^2c_6{\mathfrak f}{\mathfrak g}^2 + 16k^2c_6\Lambda {\mathfrak g}^2)\chi     \nonumber
 \\
& + (-c_5k^2{\mathfrak g}^3 + 160k^2c_6{\mathfrak f}^2{\mathfrak g} + c_5k^2{\mathfrak f}^2{\mathfrak g} + c_5k^2\Lambda {\mathfrak f}{\mathfrak g} - 16k^2c_6\Lambda {\mathfrak g}{\mathfrak f} + \frac{3}{2}k^2{\mathfrak g}\tilde{\beta})u   \nonumber
 \\
& + [32k^2c_6{\mathfrak f}^2{\mathfrak g} + 16k^2c_6\Lambda {\mathfrak g}{\mathfrak f} - c_5k^2\Lambda {\mathfrak f}{\mathfrak g} + c_5k^2{\mathfrak g}^3 + \frac{3}{2}k^2{\mathfrak g}\tilde{\beta} - c_5k^2{\mathfrak f}^2{\mathfrak g}]Q    \nonumber
 \\
& + [ic_5k^2({\mathfrak f}^2-{\mathfrak g}^2+\Lambda {\mathfrak f}) - \frac{3}{2}ik^2\tilde{\beta}]M = 0
\tag{a5}    \label{c5}
\end{align}

$(ab)$-component,  $\epsilon_{abc}k_{c}$:
\begin{align}
{\cal G}^{(\epsilon k)}  \equiv & -i(-3c_5{\mathfrak f}^2 - 3c_5\Lambda {\mathfrak f} - 2c_3\Lambda {\mathfrak f} - 2c_3{\mathfrak g}^2 - \frac{3}{2}\tilde{\beta} + 2c_3{\mathfrak f}^2 + 3c_5{\mathfrak g}^2)\rho^\prime    \nonumber
\\
& -i (4c_3{\mathfrak f}{\mathfrak g} + 12c_5{\mathfrak f}{\mathfrak g} + 96c_6{\mathfrak f}{\mathfrak g} - 48c_6\Lambda {\mathfrak g} - 2c_3\Lambda {\mathfrak g})M^\prime     \nonumber
\\
& + (-12c_5{\mathfrak f}{\mathfrak g} - 6c_5\Lambda {\mathfrak g})\Psi^\prime    \nonumber
\\
& -i (-\frac{3}{2}{\mathfrak g}\tilde{\beta} + 3c_5{\mathfrak g}^3 + 9c_5{\mathfrak f}^2{\mathfrak g} - 2c_3{\mathfrak g}^3 + 6c_3{\mathfrak f}^2{\mathfrak g} \nonumber
\\
&- 3c_5\Lambda {\mathfrak f}{\mathfrak g} - 48c_6\Lambda {\mathfrak f}{\mathfrak g} + 96c_6{\mathfrak g}{\mathfrak f}^2 - 4c_3\Lambda {\mathfrak f}{\mathfrak g})\xi     \nonumber
\\
& + (\frac{3}{2}k^2\tilde{\beta} + 2k^2c_3\Lambda {\mathfrak f} + 3k^2c_5{\mathfrak g}^2 + 2k^2c_3{\mathfrak g}^2 - 3k^2c_5{\mathfrak f}^2 - 3k^2c_5\Lambda {\mathfrak f} - 2k^2c_3{\mathfrak f}^2)u    \nonumber
\\
 & + (-96c_6{\mathfrak f}{\mathfrak g} + 2c_3\Lambda {\mathfrak g} + 48c_6\Lambda {\mathfrak g} + 6c_5\Lambda {\mathfrak g} - 4c_3{\mathfrak f}{\mathfrak g})\sigma    \nonumber
 \\
& -i (-3c_5{\mathfrak f}^3 + 48c_6{\mathfrak g}^2\Lambda + 2c_3{\mathfrak f}^3 - 6c_3{\mathfrak g}^2{\mathfrak f} - 3c_5\Lambda {\mathfrak f}^2 - 9c_5{\mathfrak g}^2{\mathfrak f} - 2c_3\Lambda {\mathfrak f}^2  \nonumber \\
& - \frac{3}{2}{\mathfrak f}\tilde{\beta} + 2c_3{\mathfrak g}^2\Lambda - 96c_6{\mathfrak f}{\mathfrak g}^2)\theta    \nonumber
\\
& + (-2c_3{\mathfrak f}^2{\mathfrak g} + 3c_5{\mathfrak f}^2{\mathfrak g} - 96c_6{\mathfrak g}{\mathfrak f}^2 + 3c_5{\mathfrak f}{\mathfrak g}\Lambda + \frac{3}{2}{\mathfrak g}\beta + 2c_3\Lambda {\mathfrak f}{\mathfrak g} - 4c_3{\mathfrak f}^2{\mathfrak g}    \nonumber
\\
& + 48c_6\Lambda {\mathfrak f}{\mathfrak g} + 2c_3\Lambda {\mathfrak f}{\mathfrak g} - 12c_5{\mathfrak f}^2{\mathfrak g} - 3c_5{\mathfrak g}^3 + 2c_3{\mathfrak g}^3)\Phi    \nonumber
\\
& + (-3c_5\Lambda {\mathfrak f}{\mathfrak g} + 3c_5{\mathfrak f}^2{\mathfrak g} - 3c_5{\mathfrak g}^3 - 2c_3{\mathfrak g}^3 - \frac{3}{2}{\mathfrak g}\tilde{\beta} \nonumber \\
& + 96c_6{\mathfrak g}{\mathfrak f}^2 + 6c_3{\mathfrak f}^2{\mathfrak g} - 4c_3\Lambda {\mathfrak f}{\mathfrak g} - 48c_6\Lambda {\mathfrak f}{\mathfrak g})\Psi    \nonumber
\\
 & -i (-6c_3{\mathfrak f}^2{\mathfrak g} + 48c_6\Lambda {\mathfrak f}{\mathfrak g} + \frac{3}{2}{\mathfrak g}\tilde{\beta} - 3c_5{\mathfrak f}^2{\mathfrak g} + 3c_5\Lambda {\mathfrak f}{\mathfrak g} \nonumber \\
 & - 96c_6{\mathfrak g}{\mathfrak f}^2 + 4c_3\Lambda {\mathfrak f}{\mathfrak g} + 3c_5{\mathfrak g}^3 + 2c_3{\mathfrak g}^3)M   \nonumber
 \\
 & -i (-6c_5{\mathfrak g}^2\Lambda + 3c_5{\mathfrak g}^2{\mathfrak f} + 2c_3\Lambda {\mathfrak f}^2 - 3c_5\Lambda {\mathfrak f}^2 - 48c_6{\mathfrak g}^2\Lambda + 96c_6{\mathfrak f}{\mathfrak g}^2 \nonumber
 \\
 & + 6c_3{\mathfrak g}^2{\mathfrak f} - 2c_3{\mathfrak g}^2\Lambda + \frac{3}{2}\tilde{\beta}{\mathfrak f} - 3c_5{\mathfrak f}^3 - 2c_3{\mathfrak f}^3)\rho = 0
\tag{a6}     \label{c6}
\end{align}

Equations obtained from \eqref{all_22} are as follows.

$(0a0)$-component:
\begin{align}
{\cal T}^{(0a0)} \equiv & -2c_5k^2\chi^\prime + i( 4c_4{\mathfrak g} + 6c_5{\mathfrak g} )\rho^\prime + 4c_3i{\mathfrak f}M^\prime + 2c_5k^2{\mathfrak g}Q \nonumber
\\
& + i(6c_5{\mathfrak g}^2 + 2k^2c_5 + 4c_3{\mathfrak f}^2 + 4c_4{\mathfrak g}^2)\xi + ( 16c_5k^2{\mathfrak g} + 4{\mathfrak g}c_4k^2 )u \nonumber
\\
& + 4c_3{\mathfrak f}\sigma + i( 8c_4{\mathfrak g}{\mathfrak f} + 18c_5{\mathfrak g}{\mathfrak f} )\theta -i ( 96\Lambda c_6{\mathfrak g} - 192c_6{\mathfrak f}{\mathfrak g} + 8c_4{\mathfrak g}{\mathfrak f} + 6c_5{\mathfrak g}{\mathfrak f} )\rho \nonumber
\\
& + ( 4c_3{\mathfrak f}^2 + 4c_4{\mathfrak g}^2 + 2k^2c_5 + 6c_5{\mathfrak g}^2 )\Phi - 2ik^2c_5{\mathfrak f}\chi  \nonumber
\\
& + ( -2c_5k^2 - 4c_4{\mathfrak g}^2 - 4c_3{\mathfrak f}^2 - 18c_5{\mathfrak g}^2 )\Psi  \nonumber
\\
& -i ( 4c_3{\mathfrak f}^2 + 18c_5{\mathfrak g}^2 - 3\tilde{\beta} + 4c_4{\mathfrak g}^2 + 2c_5k^2 )M = 0
\tag{a7}     \label{c7}
\end{align}

$(ab0)$-component:
\begin{align}
{\cal T}^{(ab0)} \equiv & -32c_6k^2u^\prime - 16c_6k^2Q^\prime - ( 96c_6{\mathfrak g} + 12c_5{\mathfrak g} )\Psi^\prime -i ( 6c_5{\mathfrak f} + 4c_4{\mathfrak f} )\rho^\prime  \nonumber
 \\
& + 4ic_3{\mathfrak g}M^\prime + 32c_6k^2{\mathfrak f}Q +i ( 96\Lambda c_6{\mathfrak g} + 6c_5{\mathfrak g}{\mathfrak f} - 192c_6{\mathfrak f}{\mathfrak g} + 8c_4{\mathfrak g}{\mathfrak f} )M   \nonumber
\\
& + ( 48c_6\Lambda {\mathfrak g} + 8c_4{\mathfrak g}{\mathfrak f} - 192c_6{\mathfrak f}{\mathfrak g} + 6c_5{\mathfrak g}{\mathfrak f} )\Psi + 32c_6k^2{\mathfrak g}\chi    \nonumber
\\
& -i ( 8c_4{\mathfrak g}{\mathfrak f} + 18c_5{\mathfrak g}{\mathfrak f} )\xi - ( -64c_6k^2{\mathfrak f} + 4{\mathfrak f}k^2c_4 + 6{\mathfrak f}k^2c_5 )u  \nonumber
\\
& + ( 96c_6{\mathfrak g} - 4c_4{\mathfrak g} )\sigma +i ( 16c_6k^2 - 4c_4{\mathfrak f}^2 - 4c_3{\mathfrak g}^2 - 6c_5{\mathfrak f}^2 )\theta  \nonumber
\\
& +i ( 32c_6k^2 + 4c_4{\mathfrak f}^2 + 3\tilde{\beta} + 18c_5{\mathfrak f}^2 + 4c_3{\mathfrak g}^2 )\rho - ( 8c_4{\mathfrak g}{\mathfrak f} + 18c_5{\mathfrak g}{\mathfrak f} - 48c_6\Lambda {\mathfrak g} )\Phi = 0
\tag{a8}     \label{c8}
\end{align}

$(0ab)$-component, $k_{a}k_{b}$:
\begin{align}
{\cal T}^{(k\otimes k)} \equiv & 3c_5\chi^{\prime\prime} + 3c_5{\mathfrak g}Q^\prime - 3c_5\Phi^\prime - 3c_5{\mathfrak g}u^\prime - (2ic_4 +3ic_5)M^\prime - 3ic_5\xi^\prime + 3c_5\Psi^\prime   \nonumber
\\
& - ( -3c_5\Lambda {\mathfrak f} + 3c_5{\mathfrak f}^2 + \frac{3}{2}\tilde{\beta} )\chi - (3ic_5{\mathfrak f} + 2ic_4{\mathfrak f})\xi   \nonumber
\\
 & - (3c_5{\mathfrak f} + 2c_4{\mathfrak f})\Phi - ( 2c_4 + 6c_5 )\sigma + 2ic_3{\mathfrak g}\rho  \nonumber
\\
& - ( 3c_5\Lambda {\mathfrak g} - 96c_6{\mathfrak f} {\mathfrak g} + 3c_5{\mathfrak f} {\mathfrak g} + 48c_6\Lambda {\mathfrak g} )u - ( -3c_5{\mathfrak f} - 2c_4{\mathfrak f} )\Psi + 2ic_4{\mathfrak g}\theta  \nonumber
\\
 & - ( -2ic_4{\mathfrak f} - 3ic_5{\mathfrak f} )M - ( -48\Lambda c_6 {\mathfrak g} + 96c_6{\mathfrak f}{\mathfrak g} - 3c_5\Lambda {\mathfrak g} - 3c_5{\mathfrak f}{\mathfrak g} )Q = 0
\tag{a9}     \label{c9}
\end{align}

$(0ab)$-component, $\delta_{ab}$:
\begin{align}
{\cal T}^{(\delta)} \equiv & c_5k^2\chi^{\prime\prime} + ( 5k^2c_5{\mathfrak g} + 32k^2c_6{\mathfrak g} )Q^\prime - c_5k^2\Phi^\prime + ( c_5k^2 + 192c_6{\mathfrak g}^2 )\Psi^\prime   \nonumber
 \\
& -ik^2c_5\xi^\prime + (-5ik^2c_5 - 2ik^2c_4)M^\prime + (64k^2c_6{\mathfrak g} + 7k^2c_5{\mathfrak g})u^\prime + ( 5ik^2c_5{\mathfrak f} + 2ik^2c_4{\mathfrak f} )M  \nonumber
\\
& +( -160k^2c_6{\mathfrak g}{\mathfrak f} + 5k^2c_5\Lambda {\mathfrak g} + 48k^2c_6\Lambda {\mathfrak g} + k^2c_5{\mathfrak f}{\mathfrak g} )Q + ( -32ik^2c_6{\mathfrak g} + 2ik^2c_4{\mathfrak g} )\theta  \nonumber
\\
& +( 48k^2c_6\Lambda {\mathfrak g} - k^2c_5{\mathfrak f}{\mathfrak g} - 224k^2c_6{\mathfrak f}{\mathfrak g} + 7k^2c_5\Lambda {\mathfrak g} )u + (-5ik^2c_5{\mathfrak f} - 2ik^2c_4{\mathfrak f})\xi  \nonumber
\\
& + ( 2k^2c_4{\mathfrak f} + 5k^2c_5{\mathfrak f} + 576c_6{\mathfrak g}^2{\mathfrak f} + 3{\mathfrak f}\tilde{\beta} - 24c_5\Lambda {\mathfrak g}^2 - 192c_6\Lambda {\mathfrak g}^2 )\Psi   \nonumber
\\
& +( - 96c_6\Lambda {\mathfrak g}^2 - 5k^2c_5{\mathfrak f} - 2k^2c_4{\mathfrak f} - 12c_5\Lambda {\mathfrak f}^2 + 3\Lambda \tilde{\beta} )\Phi   + ( -64ik^2c_6{\mathfrak g} + 2ik^2c_3{\mathfrak g} )\rho  \nonumber
\\
& +( 2k^2c_3 - 3\tilde{\beta} - 192c_6{\mathfrak g}^2 )\sigma + ( -k^2c_5{\mathfrak f}^2 - \frac{3}{2}k^2\tilde{\beta} + k^2c_5\Lambda {\mathfrak f} - 64k^2c_6{\mathfrak g}^2 )\chi = 0
\tag{a10}     \label{c10}
\end{align}

$(0ab)$-component, $\epsilon_{abc}k_{c}$:
\begin{align}
{\cal T}^{(\epsilon k)} \equiv & i(2c_4+3c_5)\rho^{\prime\prime} - 3ic_5{\mathfrak g}M^\prime - 16k^2c_6Q^\prime +i ( 2c_4{\mathfrak f} + 3c_5{\mathfrak f} )\theta^\prime +i ( 2c_4{\mathfrak g}+3c_5{\mathfrak g} )\xi^\prime   \nonumber
\\
& + ( 3k^2c_5 + 2k^2c_4 - 32k^2c_6 )u^\prime + ( 3c_5{\mathfrak g} + 2c_4{\mathfrak g} )\Phi^\prime + ( -2c_4{\mathfrak g} - 9c_5{\mathfrak g} - 96c_6{\mathfrak g} )\Psi^\prime   \nonumber
\\
& -i ( -3c_5{\mathfrak g}{\mathfrak f} - 48c_6\Lambda {\mathfrak g} - 4c_4{\mathfrak g}{\mathfrak f} + 96c_6{\mathfrak g}{\mathfrak f} - 2c_4\Lambda {\mathfrak g} - 3c_5\Lambda {\mathfrak g} )\xi    \nonumber
\\
 & + ( 2{\mathfrak f}k^2c_4 + 64k^2c_6{\mathfrak f} + 3k^2{\mathfrak f}c_5 )u + ( 96c_6{\mathfrak g} + 6c_5{\mathfrak g} + 2c_4{\mathfrak g} )\sigma  \nonumber
 \\
& -i ( -9c_5{\mathfrak f}^2 - 2c_4\Lambda {\mathfrak f} + 6c_5{\mathfrak g}^2 - 2c_4{\mathfrak f}^2 - 16c_6k^2 - \frac{3}{2}\tilde{\beta} - 3c_5\Lambda {\mathfrak f} + 2c_4{\mathfrak g}^2 )\theta   \nonumber
\\
 & -i ( -2c_4{\mathfrak g}^2 - 32c_6k^2 + 9c_5\Lambda {\mathfrak f} - 6c_5{\mathfrak g}^2 - \frac{3}{2}\tilde{\beta} + 2c_4\Lambda {\mathfrak f} + 2c_4{\mathfrak f}^2 + 3c_5{\mathfrak f}^2 )\rho  \nonumber
\\
& + ( 3c_5\Lambda {\mathfrak g} + 2c_4{\mathfrak f}{\mathfrak g} + 2c_4\Lambda {\mathfrak g} + 2c_4{\mathfrak g}{\mathfrak f} + 48c_6\Lambda {\mathfrak g} + 3c_5{\mathfrak f}{\mathfrak g} )\Phi   \nonumber
\\
& + ( -4c_4{\mathfrak g}{\mathfrak f} - 2c_4\Lambda {\mathfrak g} - 192c_6{\mathfrak f}{\mathfrak g} + 48c_6\Lambda {\mathfrak g} - 9c_5{\mathfrak f}{\mathfrak g} - 3c_5\Lambda {\mathfrak g} )\Psi   \nonumber
\\
& + 32c_6k^2{\mathfrak g}\chi -i ( 4c_4{\mathfrak g}{\mathfrak f} + 2c_4\Lambda {\mathfrak g} + 3c_5\Lambda {\mathfrak g} + 9c_5{\mathfrak g}{\mathfrak f} + 96c_6{\mathfrak f}{\mathfrak g} - 48c_6\Lambda {\mathfrak g} )M  \nonumber
\\
& + 32k^2c_6{\mathfrak f}Q = 0
\tag{a11}     \label{c11}
\end{align}

$(abc)$-component, $\epsilon_{abd}k_{c}k_{d}$:
\begin{align}
{\cal T}^{(\epsilon k \otimes k)} \equiv & -(3ic_5+2ic_4)\rho^\prime + 3c_5{\mathfrak g}\chi^\prime + 2ic_3{\mathfrak g}\xi - (3c_5{\mathfrak g}^2 - \frac{3}{2}\tilde{\beta} + 3k^2c_5 + 2k^2c_4 + 6c_5\Lambda {\mathfrak f})u  \nonumber
\\
 & - (2ic_4{\mathfrak f}+3ic_5{\mathfrak f})\theta + (2ic_4{\mathfrak f} + 3ic_5{\mathfrak f})\rho + 2c_3{\mathfrak g}\Phi - (48c_6\Lambda {\mathfrak g} - 96c_6{\mathfrak f}{\mathfrak g} + 3c_5{\mathfrak g}{\mathfrak f})\chi   \nonumber
 \\
 & + 2c_4{\mathfrak g}\Psi + 2ic_4{\mathfrak g}M - (\frac{3}{2}\tilde{\beta} - 6c_5\Lambda {\mathfrak f} - 3c_5{\mathfrak g}^2)Q = 0
\tag{a12}     \label{c12}
\end{align}

$(abc)$-component, $\delta_{ac}k_{b} - \delta_{ab}k_{c}$:
\begin{align}
{\cal T}^{(\delta\otimes k)} \equiv & 2ic_3M^{\prime\prime} - 2c_3{\mathfrak f}\Psi^\prime + 2c_3{\mathfrak f}\Phi^\prime + 3ic_5{\mathfrak g}\rho^\prime - 2ic_3{\mathfrak g}\theta^\prime + 2ic_3{\mathfrak f}\xi^\prime + 2c_3\sigma^\prime - k^2c_5\chi^\prime   \nonumber
\\
& -i (-2c_3{\mathfrak f}^2 + \frac{3}{2}\tilde{\beta} - k^2c_5 - 2c_3\Lambda {\mathfrak f} - 2c_4{\mathfrak g}^2 - 9c_5{\mathfrak g}^2 )\xi + (2{\mathfrak g}c_4k^2 + 2{\mathfrak g}c_5k^2)u   \nonumber
 \\
& + 2c_3{\mathfrak f}\sigma -i ( 2c_3\Lambda {\mathfrak g} - 3c_5{\mathfrak f}{\mathfrak g} + 96c_6 {\mathfrak f}{\mathfrak g} - 48c_6\Lambda {\mathfrak g} - 4c_4{\mathfrak g}{\mathfrak f} )\theta   \nonumber
 \\
 & -i (-48c_6\Lambda {\mathfrak g} + 4c_4{\mathfrak g}{\mathfrak f} + 9c_5 {\mathfrak g}{\mathfrak f} + 2c_4\Lambda {\mathfrak g} + 96c_6 {\mathfrak f}{\mathfrak g})\rho \nonumber \\
 & + (2c_3\Lambda {\mathfrak f} + 2c_3{\mathfrak f}^2 - 2c_3{\mathfrak g}^2 + 3c_5{\mathfrak g}^2 + c_5k^2)\Phi  \nonumber
  \\
& + ( -2c_3\Lambda {\mathfrak f} - 2c_4{\mathfrak g}^2 - 2c_3{\mathfrak f}^2 - 3c_5{\mathfrak g}^2 - k^2c_5 )\Psi -  c_5k^2{\mathfrak f}\chi  \nonumber
\\
& -i ( k^2c_5 + \frac{3}{2}\tilde{\beta} + 2c_4{\mathfrak g}^2 + 2c_3{\mathfrak f}^2 + 3c_5{\mathfrak g}^2 + 2c_3\Lambda {\mathfrak f} )M + c_5k^2{\mathfrak g}Q = 0
\tag{a13}     \label{c13}
\end{align}

$(abc)$-component, $\epsilon_{abc}$:
\begin{align}
{\cal T}^{(\epsilon)} \equiv & -32c_6k^2u^{\prime\prime} - 16c_6k^2Q^{\prime\prime} + (-96c_6{\mathfrak g} - 12c_5{\mathfrak g})\Psi^{\prime\prime} + (5k^2c_5{\mathfrak g} + 32k^2c_6{\mathfrak g})\chi^\prime   \nonumber
 \\
& + ( 32ic_6k^2 - 2ik^2c_4 - 3ik^2c_5 )\rho^\prime + ( -48c_6\Lambda {\mathfrak g} - 384c_6{\mathfrak f}{\mathfrak g} )\Psi^\prime + 48c_6\Lambda {\mathfrak g}\Phi^\prime + 16ik^2c_6\theta^\prime  \nonumber
 \\
 &  + (12c_5{\mathfrak g}+96c_6{\mathfrak g})\sigma^\prime   + (-8ik^2c_5{\mathfrak g} - 2ik^2c_4{\mathfrak g})\xi   + ( 2ik^2c_4{\mathfrak g} + 2ik^2c_5{\mathfrak g} )M    \nonumber
 \\
& + ( 32ik^2c_6{\mathfrak f} - 3ik^2c_5{\mathfrak f} - 2ik^2c_4{\mathfrak f} )\theta + ( 160k^2c_6{\mathfrak g}{\mathfrak f} - k^2c_5{\mathfrak f}{\mathfrak g} - 16k^2\Lambda {\mathfrak g} c_6 )\chi \nonumber
\\
& + ( -3k^4c_5 - 2k^4c_4 + 64k^2c_6\Lambda {\mathfrak f} - \frac{3}{2}k^2\tilde{\beta} - k^2c_5{\mathfrak g}^2 + 6k^2c_5\Lambda {\mathfrak f} + 128k^2c_6{\mathfrak f}^2 )u  \nonumber
 \\
& + (-2k^2c_4{\mathfrak g} - 3k^2c_5{\mathfrak g} + 96c_6\Lambda^2 {\mathfrak g} - 12c_5\Lambda {\mathfrak f} {\mathfrak g} - 5k^2c_5{\mathfrak g} + 96c_6\Lambda {\mathfrak f} {\mathfrak g})\Phi  \nonumber
\\
& + ( 32k^2c_6\Lambda {\mathfrak f} + 64k^2c_6{\mathfrak f}^2 + k^2c_5{\mathfrak g}^2 +6k^2c_5\Lambda {\mathfrak f} - \frac{3}{2}k^2\tilde{\beta} )Q + 384c_6{\mathfrak f}{\mathfrak g}\sigma \nonumber
\\
& + ( 2k^2c_5{\mathfrak g} - 192c_6\Lambda {\mathfrak f} {\mathfrak g} + 3{\mathfrak g}\tilde{\beta} - 576c_6{\mathfrak f}^2{\mathfrak g} + 96c_6\Lambda^2{\mathfrak g} + 2k^2c_4{\mathfrak g} - 24c_5\Lambda {\mathfrak f} {\mathfrak g} )\Psi  \nonumber
 \\
& + ( 3ik^2c_5{\mathfrak f} + 64ik^2c_6{\mathfrak f} + 2ik^2c_4{\mathfrak f} )\rho = 0
\tag{a14}      \label{c14}
\end{align}

Note that eqs.~\eqref{c5}, \eqref{c9}, \eqref{c10}, \eqref{c11}, \eqref{c13} and \eqref{c14} are second order, while eqs.~\eqref{c1}-\eqref{c4}, \eqref{c6}-\eqref{c8} and \eqref{c12} are first order.

\section*{Appendix B} \label{B}
Here we give explicit forms of the first order equations which replace the eqs. \eqref{c11} and \eqref{c12}:
\begin{align}
{\cal D}^{(1)} & \equiv   4ic_3k^2M^\prime      + ( 12c_5{\mathfrak g}k^2 + 96c_6{\mathfrak g}k^2)Q^\prime + 576c_6{\mathfrak g}^2\Psi^\prime + 2( 12c_5{\mathfrak g}k^2 + 96c_6{\mathfrak g}k^2)u^\prime   \nonumber
\\
& + (  4c_3k^2 - 9\tilde{\beta}- 576c_6{\mathfrak g}^2)\sigma  + 4ik^2c_3{\mathfrak f}\xi + ( -192ik^2c_6{\mathfrak g} + 4ik^2c_3{\mathfrak g} )\rho \nonumber \\
&+ ( 12k^2c_5\Lambda {\mathfrak g} + 96c_6k^2\Lambda {\mathfrak g} - 384c_6k^2{\mathfrak g}{\mathfrak f} )Q + 2( 12k^2c_5\Lambda {\mathfrak g} \nonumber \\
& -  4ik^2c_3{\mathfrak f}M + 96c_6k^2\Lambda {\mathfrak g} - 384c_6k^2{\mathfrak g}{\mathfrak f} )u \nonumber \\
&+ ( -96c_6ik^2{\mathfrak g} + 4ik^2c_4{\mathfrak g} )\theta    + (-3k^2\tilde{\beta}  - 192k^2c_6{\mathfrak g}^2)\chi \nonumber
\\
&+ (  4k^2c_3{\mathfrak f} - 36c_5\Lambda {\mathfrak f}^2 - 288c_6\Lambda {\mathfrak g}^2 +9\tbeta \Lambda )\Phi \nonumber \\
& + [ -72c_5\Lambda {\mathfrak g}^2 -  4k^2c_3{\mathfrak f} + 9\tilde{\beta}({\mathfrak f}+\Lambda) - 576c_6{\mathfrak g}^2(\Lambda - 3{\mathfrak f}) ]\Psi  = 0  \;, \tag{b1} \label{d1}
\end{align}

\begin{align}
{\cal D}^{(2)} & \equiv [ 3\tilde{\beta}-12c_5\Lambda {\mathfrak f} ]u^\prime + [ -3\tilde{\beta} + 12c_5\Lambda {\mathfrak f} ]Q^\prime + [ 6c_5\Lambda {\mathfrak g} - 96c_6{\mathfrak g}(\Lambda-2{\mathfrak f}) ]\chi^\prime  \nonumber
\\
& + 8(c_4-c_3)i{\mathfrak g}M^\prime   + 4(c_4-c_3)i{\mathfrak f}\rho^\prime + [ 3i\tilde{\beta} - 8ic_3({\mathfrak f}^2-{\mathfrak g}^2) - 4(c_4-c_3)i{\mathfrak g}^2 ]\theta \nonumber
\\
&  + 8(c_4-c_3){\mathfrak g}\sigma   + [ 96ic_6{\mathfrak g}(\Lambda-2{\mathfrak f}) - 6ic_5\Lambda {\mathfrak g} + 12(c_4-c_3)i{\mathfrak f}{\mathfrak g} ]\xi \nonumber
\\
& + [ -96c_6{\mathfrak g}i(\Lambda-2{\mathfrak f}) - 6ic_5\Lambda {\mathfrak g} - 12i(c_4-c_3){\mathfrak f}{\mathfrak g} ]M  \nonumber
\\
& + [ -3i\tilde{\beta} - 12ic_5\Lambda {\mathfrak f} + 16ic_3({\mathfrak f}^2-{\mathfrak g}^2) + 4(c_4-c_3)i{\mathfrak f}^2 ]\rho \nonumber \\
& + [ -12(c_4-c_3){\mathfrak f}{\mathfrak g} - 6c_5\Lambda {\mathfrak g} ]\Psi + [ 12(c_4-c_3){\mathfrak f}{\mathfrak g} - 6c_5\Lambda {\mathfrak g} ]\Phi \nonumber
\\
& + [ 3\tilde{\beta}{\mathfrak g} - 18c_5\Lambda {\mathfrak f}{\mathfrak g} + 192c_6{\mathfrak f}^2{\mathfrak g} - 192c_6\Lambda^2{\mathfrak g} + 288c_6\Lambda {\mathfrak f}{\mathfrak g} ]\chi \nonumber
\\
& + [ 96c_6{\mathfrak g}^2(\Lambda-2{\mathfrak f}) + 6\Lambda \tilde{\beta} + 3{\mathfrak f}\tilde{\beta} +  4(c_4-c_3){\mathfrak f}k^2 - 6c_5\Lambda {\mathfrak g}^2 - 12c_5\Lambda {\mathfrak f}^2 - 24c_5\Lambda^2{\mathfrak f} ]u \nonumber
\\
& + [ -3{\mathfrak f}\tilde{\beta} - 6\Lambda \tilde{\beta} - 96c_6{\mathfrak g}^2(\Lambda-2{\mathfrak f}) + 6c_5\Lambda {\mathfrak g}^2 + 12c_5\Lambda {\mathfrak f}^2 + 24c_5\Lambda^2 {\mathfrak f} ]Q = 0 \;.  \tag{b2} \label{d2}
\end{align}
These equations are obviously first order.

\end{document}